\documentclass[reprint,superscriptaddress,amsmath,amssymb,prl,showpacs,floatfix]{revtex4-2}
\usepackage{cases}
\usepackage{amsmath}
\usepackage{amssymb}
\usepackage{amsfonts}
\usepackage{amssymb}
\usepackage{dcolumn}
\usepackage{bm}
\usepackage{natbib}
\usepackage{hyperref}
\hypersetup{colorlinks=true,citecolor=red}
\usepackage{cleveref}
\usepackage{graphicx}
\usepackage{upgreek}
\usepackage{subfigure}
\usepackage{color}
\usepackage{xcolor}
\usepackage{textcomp}
\usepackage{siunitx}
\usepackage{physics}
\usepackage{booktabs}
\usepackage[normalem]{ulem}
\UseRawInputEncoding
\usepackage[columnwise]{lineno}
\usepackage[integrals]{wasysym}
\DeclareSIUnit\cps{\mathrm{cps}}
\DeclareSIUnit\Molar{\textsc{M}}
\DeclareSIUnit\rpm{\mathrm{rpm}}
\DeclareSIUnit\gauss{G}

\newcolumntype{x}{>{$}X<{$}}
\begin{document}
	
\title{Removing inhomogeneous broadening in cross-relaxation spectra via hole burning}

\author{Fei Kong}
\altaffiliation{These authors contributed equally to this work.}
\affiliation{Laboratory of Spin Magnetic Resonance, School of Physical Sciences, Anhui Province Key Laboratory of Scientific Instrument Development and Application, University of Science and Technology of China, Hefei 230026, China}
\affiliation{Hefei National Research Center for Physical Sciences at the Microscale, Hefei 230026, China}
\affiliation{Hefei National Laboratory, University of Science and Technology of China, Hefei 230088, China}

\author{Zhehua Huang}
\altaffiliation{These authors contributed equally to this work.}
\affiliation{Laboratory of Spin Magnetic Resonance, School of Physical Sciences, Anhui Province Key Laboratory of Scientific Instrument Development and Application, University of Science and Technology of China, Hefei 230026, China}
\affiliation{Hefei National Research Center for Physical Sciences at the Microscale, Hefei 230026, China}

\author{Zhengze Zhao}
\affiliation{Laboratory of Spin Magnetic Resonance, School of Physical Sciences, Anhui Province Key Laboratory of Scientific Instrument Development and Application, University of Science and Technology of China, Hefei 230026, China}
\affiliation{Hefei National Research Center for Physical Sciences at the Microscale, Hefei 230026, China}

\author{Pengju Zhao}
\affiliation{Laboratory of Spin Magnetic Resonance, School of Physical Sciences, Anhui Province Key Laboratory of Scientific Instrument Development and Application, University of Science and Technology of China, Hefei 230026, China}

\author{Zhecheng Wang}
\affiliation{Laboratory of Spin Magnetic Resonance, School of Physical Sciences, Anhui Province Key Laboratory of Scientific Instrument Development and Application, University of Science and Technology of China, Hefei 230026, China}
\affiliation{School of Biomedical Engineering and Suzhou Institute for Advanced Research, University of Science and Technology of China, Suzhou 215123, China}

\author{Ya Wang}
\affiliation{Laboratory of Spin Magnetic Resonance, School of Physical Sciences, Anhui Province Key Laboratory of Scientific Instrument Development and Application, University of Science and Technology of China, Hefei 230026, China}
\affiliation{Hefei National Research Center for Physical Sciences at the Microscale, Hefei 230026, China}
\affiliation{Hefei National Laboratory, University of Science and Technology of China, Hefei 230088, China}

\author{Fazhan Shi}
\email{fzshi@ustc.edu.cn}
\affiliation{Laboratory of Spin Magnetic Resonance, School of Physical Sciences, Anhui Province Key Laboratory of Scientific Instrument Development and Application, University of Science and Technology of China, Hefei 230026, China}
\affiliation{Hefei National Research Center for Physical Sciences at the Microscale, Hefei 230026, China}
\affiliation{Hefei National Laboratory, University of Science and Technology of China, Hefei 230088, China}
\affiliation{School of Biomedical Engineering and Suzhou Institute for Advanced Research, University of Science and Technology of China, Suzhou 215123, China}

\begin{abstract}
	Quantum relaxometry based on nitrogen-vacancy (NV) centers in diamond is an easy-to-use technique that detects magnetic noise by measuring the longitudinal relaxation of NV centers. It is favored in chemical and biological applications due to the robustness against imperfect spin control. By tuning the energy level, NV relaxometry can detect magnetic noise generated by other spins and obtain magnetic resonance spectra through cross relaxation. However, the inhomogeneous broadening of NV centers greatly limits the spectral resolution of cross-relaxation spectra. Here we demonstrate a hole-burning technique to remove the inhomogeneous broadening. We utilize a weak pump field to deplete specific NV centers during the relaxation measurement, which contribute a reverse signal in the cross-relaxation spectra with a much narrower linewidth. Our method retains the convenience and sensitivity of NV relaxometry, opening up new avenues for high-resolution magnetic resonance spectroscopy in complex scenarios.
\end{abstract}

\maketitle


Magnetic resonance spectroscopy is a widely used analytical tool for composition quantification and structure determination \cite{Lambert2019,Borbat2001}. Quantum sensors \cite{Degen2017}, especially nitrogen-vacancy (NV) centers in diamond \cite{Doherty2013}, can extend the magnetic resonance technique to new areas beyond the reach of conventional means, such as detection at the single-molecule scale \cite{Du2024} or at interfaces \cite{Liu2022}. The NV centers are usually used as atomic-scale magnetometers to coherently detect the magnetic field generated by other electron or nuclear spins. This magnetometry-based magnetic resonance uses sophisticated spin-control pulses to filter out the fields generated by environmental spins \cite{Staudacher2013,Mamin2013,Grinolds2014}, and thus can identify target spins with high spectral resolution \cite{Staudacher2013,Schmitt2017,Boss2017,Glenn2018}. Another less commonly used technique is NV relaxometry, which has minimal requirements on spin controls \cite{Sushkov2014,Hall2016,Simpson2017,Qin2023}. It can be easily applied on flexible nanodiamonds (NDs) or large ensemble of NV centers \cite{Kaufmann2013,Steinert2013,Qin2023}, and is thus more favored for applications in complex scenarios \cite{Mzyk2022}, for instance, detecting cellular activities \cite{Nie2021,Sharmin2021,Tian2022} and monitoring chemical reactions \cite{Rendler2017,Simpson2017,Barton2020,Martinez2020}. As a magnetic resonance tool, the NV relaxometry has demonstrated comparable sensitivity to the NV magnetometry \cite{Wood2017}, but its spectral resolution is much worse \cite{Hall2016,Simpson2017,Wood2017,Kong2018,Kong2020}. To unambiguously identify the signal sources in these complex scenarios, it is necessary to improve the spectral resolution of NV relaxometry.

For NV relaxometry, the magnetic resonance signal arises from an increase in the NV relaxation rate ($\Gamma_1=1/T_1$) when the energy splittings of the NV and the target match. By sweeping the detuning, a cross-relaxation resonance spectrum can be acquired \cite{Hall2016,Simpson2017,Wood2017,Qin2023}. The inhomogeneous broadening of the NV centers, determined by the dephasing rate ($\Gamma_2^*=1/T_2^{*}$), hinders the energy matching, and thus limits the spectral resolution. For NV magnetometry, this limitation can be released to the decoherence rate ($\Gamma_2=1/T_2$) by the dynamical decoupling technique \cite{Staudacher2013}, and can be further improved to $\Gamma_1$ or even better by correlation or heterodyne detection \cite{Laraoui2013,Schmitt2017,Boss2017}. However, those techniques are not compatible with the cross-relaxation method. Spectral hole burning, a well-established technique in optics \cite{Moerner1988}, can also remove the inhomegeneous broadening. It usually requires a pump field at a frequency within the inhomogeneous linewidth to selectively deplete specific emitters whose transitions are resonant with the pump field, creating a ``hole" in the spectral profile when another probe field is swept across the resonance frequency \cite{Moerner1988}. This technique has been employed to acquire high-resolution spectra of NV itself \cite{Kehayias2014,Ella2019}, showing compatibility with NV relaxometry. 

In this work, we show that the hole-burning technique can also been employed to remove the inhomogeneous broadening in cross-relaxation spectra of other spins, where an artificial radiofrequency (RF) field serves as the pump field, while the NV-target coupling serves as the probe field. We perform the experimental demonstration on detection of nuclear spins on diamond surface. By tuning the RF strength, the cross-relaxation nuclear magnetic resonance (NMR) spectrum can be narrowed without sacrifice of sensitivity. As an intuitive example, we show the resonance lines of $^{1}$H and $^{19}$F can be clearly resolved in the hole burning spectrum, but are indistinguishable in the normal cross-relaxation spectrum. We also develop a simple theoretical model to quantitatively characterize the hole-burning spectra, which shows good agreement with the experiment.

A common cross-relaxation NMR model consists of a sensor and a target. As shown in Fig.~1a, the NV sensor has a energy splitting of $\omega$ between $|0\rangle$ and $|1\rangle$ with an additional shift of $\Delta$ induced by inhomogeneity. The spectral line shape of NV sensors is normally characterized by a Gaussian distribution $G(\Delta;\sqrt{2}\Gamma_2^*)$ with a full width at half maximum (FWHM) linewidth of $4\sqrt{\ln 2}\Gamma_2^*$. The target nuclear spin has a energy splitting of $\nu$ between $\left|\uparrow\right\rangle$ and $\left|\downarrow\right\rangle$ with a negligible linewidth comparing with the sensor. The NV sensor is initially polarized to a ``cold" state $|0\rangle$ by 532 nm laser, while the target spin stays in a thermal equilibrium state $0.5\left|\uparrow \rangle\langle\uparrow\right|+0.5\left|\downarrow\rangle\langle\downarrow\right|$. After turning off the laser, the NV state will relax to the thermal equilibrium state with a rate of $\Gamma_1$ (Fig.~1b). When $\omega\approx\nu$, the target spin will ``heating" the NV sensor, and open an extra relaxation channel with rate of
\begin{equation}
	\Gamma_{\text{d}}=\frac{d^2 \Gamma_2}{\Gamma_2^2 + (\omega-\nu+\Delta)^2},
\end{equation}
where $d$ is the dipole-dipole coupling between the sensor and the target. The cross-relaxation spectrum is obtained by sweeping $\omega$ and measuring the population of $|0\rangle$, which is given by
\begin{equation}
	\begin{split}
	& S(\omega) =\int\frac{1}{2}\left[1+e^{-(\Gamma_1+\Gamma_{\text{d}}) t}\right]\cdot G(\Delta)d\Delta \\
	& \approx \frac{1}{2}+\frac{e^{-\Gamma_1 t}}{2} \int[1-\frac{(1-e^{-\frac{d^2 t}{\Gamma_2}})(\alpha \Gamma_2)^2}{(\alpha\Gamma_2)^2 + (\omega-\nu+\Delta)^2}]G(\Delta)d\Delta \\
	& = \frac{1}{2}+\frac{e^{-\Gamma_1 t}}{2}-C\cdot L(\omega-\nu;\alpha\Gamma_2)*G(\omega;\sqrt{2}\Gamma_2^*),
	\end{split}
\end{equation}
where we use a broadened Lorentz function to approximate the exponential-Lorentz function and $\alpha\approx1+0.3d^2 t/\Gamma_2$ (see SI Note 1). The spectrum is a convolution of the NV line shape and a Lorentz function, as shown in Fig.~1c, and dominated by the former because $\Gamma_2 \ll \Gamma_2^*$. Therefore, the cross-relaxation NMR spectrum usually has a line shape similar to the NV sensor with signal contrast of
\begin{equation}
	C_{\text{T}}=\frac{\sqrt{\pi}\alpha\Gamma_2}{4\Gamma_2^*}e^{-\Gamma_1 t}(1-e^{-\frac{d^2 t}{\Gamma_2}}),
\end{equation}

\begin{figure}
	\centering \includegraphics[width=1\columnwidth]{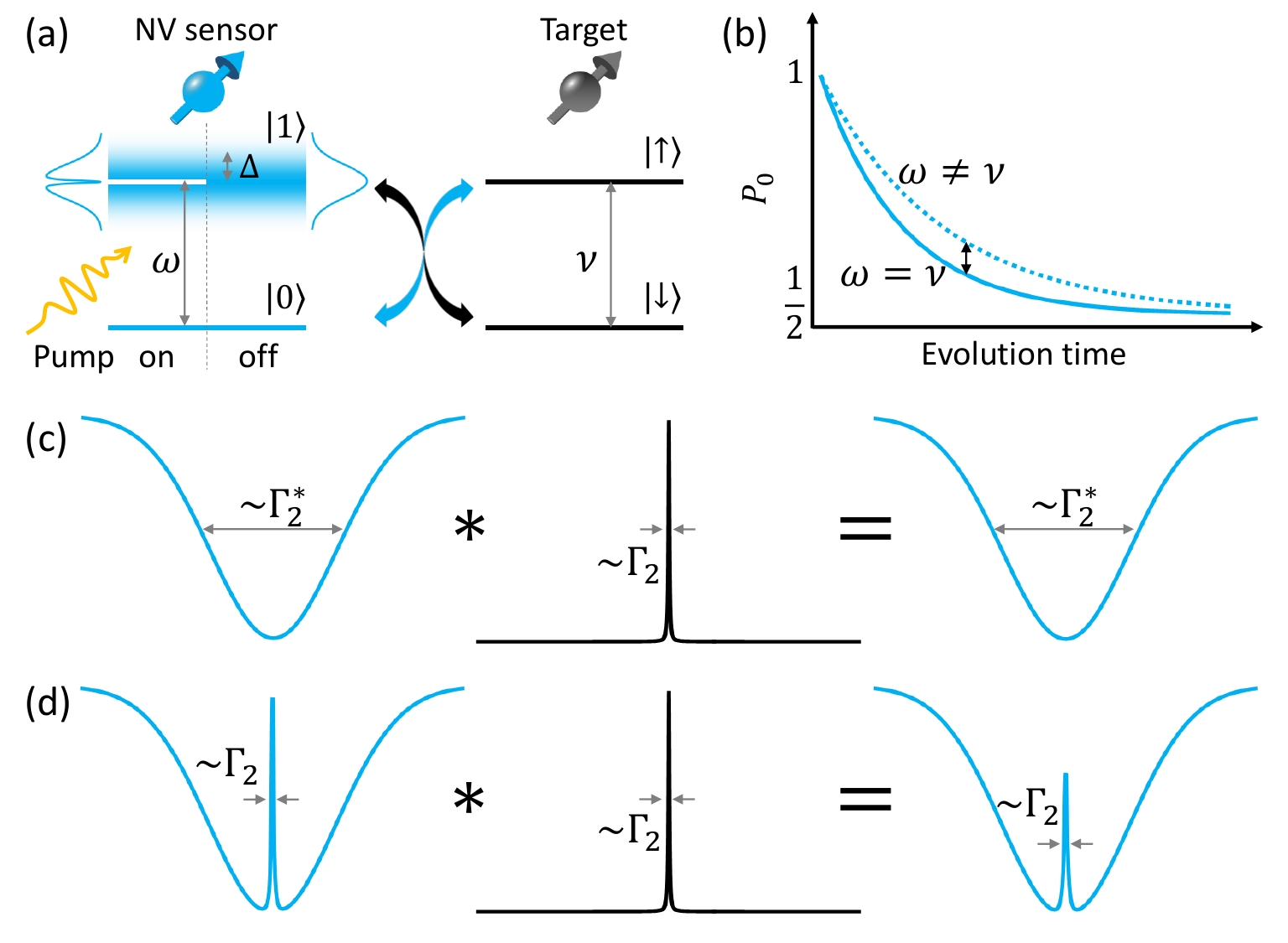} \protect\caption{
		{Hole burning in cross-relaxation spectra.}
		(a) Cross-relaxation model. The energy level of the NV sensor is broadened by inhomogeneous noise $\Delta$, whose distribution gives the NV line shape. A weak pump field will heat a specific NV from its initial $|0\rangle$ state to the thermal equilibrium state, and thus destroy its sensing ability. 
		(b) Relaxation measurement. The population of $|0\rangle$ ($P_0$) has a exponential decay with a rate of $\Gamma_1$. When the energy levels of the NV sensor and the target match, the decay rate will be faster, giving a resonant cross-relaxation signal.
		(c)(d) Cross-relaxation spectra with pump off (c) and on (d). The measured spectrum is a convolution of the NV line shape and a Lorentz function. The latter corresponds to a cross-relaxation spectrum without inhomogeneous broadening. (d) When part of the NV sensor is depleted, the NV line shape shows a burned hole. 
	}
	\label{scheme}
\end{figure}

If applying an RF field with frequency of $f$, the relaxation rate will increase by
\begin{equation}
	\Gamma_{\text{b}}=\frac{b^2 \Gamma_2}{\Gamma_2^2 + (\omega-f+\Delta)^2},
\end{equation}
where $b=\gamma_{\text{NV}}B_{\text{RF}}$ is the driving strength with $\gamma_{\text{NV}}$ and $B_{\text{RF}}$ being the gyromagnetic ratio of NV center and RF field amplitude, respectively. As long as the RF frequency $f$ sweeps in sync with the NV frequency $\omega$, the additional relaxation $\exp(-\Gamma_{\text{b}}t)$ can be absorbed to the NV line shape as
\begin{equation}
	\begin{split}
	G^{'}(\Delta)& =\frac{1}{2\sqrt{\pi}\Gamma_2^*}e^{-\left[\frac{\Delta^2}{4\Gamma_2^{*2}}+\frac{b^2 \Gamma_2 t}{\Gamma_2^2 + \Delta^2}\right]} \\
	& \approx \frac{1}{2\sqrt{\pi}\Gamma_2^*}e^{-\frac{\Delta^2}{4\Gamma_2^{*2}}}-\frac{(1-e^{-\frac{b^2 t}{\Gamma_2}})(\beta \Gamma_2)^2}{2\sqrt{\pi}\Gamma_2^* \left[ (\beta\Gamma_2)^2 + \Delta^2 \right]},
	\end{split}
\end{equation}
where $\beta\approx1+0.3b^2 t/\Gamma_2$. Here we use a similar approximation as Eq.~2 and assume $G(\Delta)\approx G(0)$ around the Lorentz peak. As shown in Fig.~1d, the NV line shape now has a burned hole, which means that part of the NV sensor is depleted and thus loses the ability to sensing. Similar to Eq.~2, the hole-burning spectrum can be calculated as
\begin{equation}
	S_{\text{HB}}(\omega) \approx S(\omega)-C_{\text{RF}}+C_{\text{HB}}\cdot L_0(\omega-\nu;(\alpha+\beta)\Gamma_2),
\end{equation}
where 
\begin{equation}
	C_{\text{RF}}=\frac{\sqrt{\pi}\beta\Gamma_2}{4\Gamma_2^*}e^{-\Gamma_1 t}(1-e^{-\frac{b^2 t}{\Gamma_2}})
\end{equation}
is the additional population loss induced by the RF field,
\begin{equation}
	C_{\text{HB}}=\frac{\sqrt{\pi}\alpha\beta\Gamma_2}{4(\alpha+\beta)\Gamma_2^*}e^{-\Gamma_1 t}(1-e^{-\frac{d^2 t}{\Gamma_2}})(1-e^{-\frac{b^2 t}{\Gamma_2}})
\end{equation}
is the hole-burning signal contrast, and $L_0$ is a normalized Lorentz function with unit height rather than unit area. Therefore, the burned hole has a linewidth of
\begin{equation}
	\text{FWHM}_{\text{HB}}=4\Gamma_2 + 0.6	d^2 t + 0.6 b^2 t,
\end{equation}
which is determined by $\Gamma_2$ rather than $\Gamma_2^*$.

\begin{figure}[htbp]
	\centering \includegraphics[width=1\columnwidth]{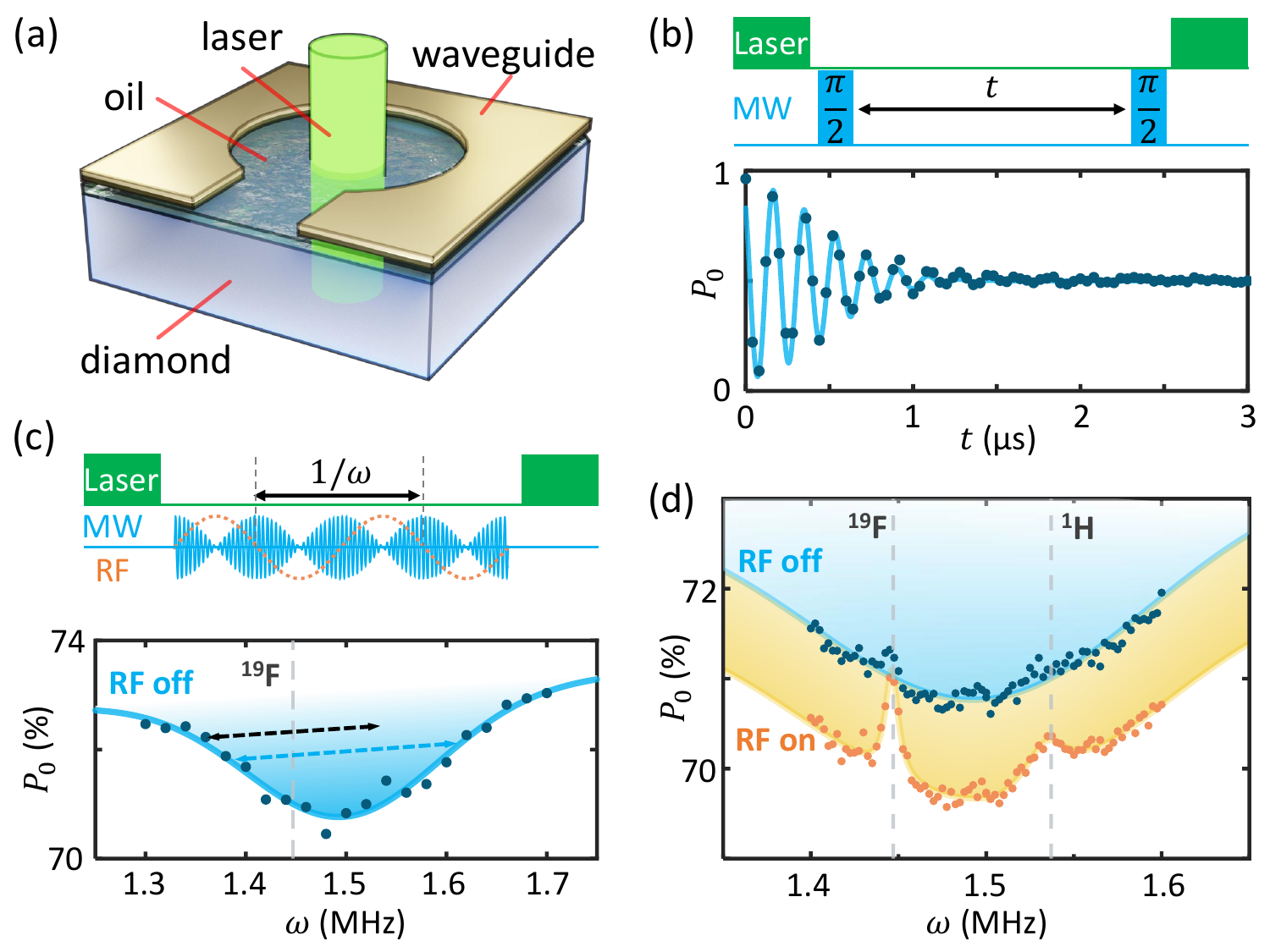} \protect\caption{
		{Experimental demonstration of hole-burning spectra.}
		(a) Schematic diagram of the key component of the experimental setup. It is placed on the sample stage of a wide-field microscope. The NV centers are located $\sim$10 nm below the diamond-sample interface.
		(b) Ramsey measurement. The points are experimental results, where $P_0$ is given by the fluorescence intensity normalized by Rabi oscillation. The line is a cosine fit with damping of $\exp[-(t/t_c)^2]$, where the fitting result gives $t_c= 0.63(2)$ \textmu{}s.
		(c) Measurement of cross-relaxation spectra. The data points are measured using the pulse sequence shown in the upper panel with pump RF turning off, $\kappa=2.0$, and evolution time of 0.2 ms. The sequence is repeated 0.5 million times. The solid line is a Gaussian fit with a sloping baseline. The gray vertical dashed line marks the resonance frequency of $^{19}$F nuclear spins with an estimated linewidth shown by the black dashed arrow. The blue dashed arrow shows the fitting linewidth.
		(d) Measurement of hole-burning spectra. The measurement is similar to (c), but with a finer frequency sweep and pump RF turning on/off. The sequence is repeated 1 million times. The strength of RF field is 1.88 kHz. The blue solid line is identical to (c), while the orange solid line is a two-peak Lorentz fit with baseline given by (c). The two vertical dashed lines mark the resonance frequencies of $^{19}$F and $^1$H nuclear spins. 
	}
	\label{demo}
\end{figure}

To demonstrate the high-resolution ability of hole-burning spectra, we use an ensemble of NV centers as the sensor whose linewidth is dominated by the inhomogeneity of local fields. The experimental setup is a homemade wide-field microscope, which is the same as ref.\cite{Huang2025}. Figure~2a only shows the key component, which consists of a diamond with samples on the surface and a microwave(MW) waveguide. The diamond is 100-oriented with an $^{12}$C isotropic purified layer. The NV centers are created by implantation of $^{15}$N$^+$ ions with energy of 5 keV and dose of $2\times 10^{12}$ cm$^{-2}$ and subsequent annealing. The sample is a drop of Fomblin YH VAC 140/13 oil containing $^{19}$F nuclear spins with density of $\sim$40 nm$^{-3}$ \cite{DeVience2015}. The coplanar waveguide fabricated on a glass substrate has an $\Omega$-shape transparent region with a diameter of 50 \textmu{}m. The microscope's field of view is $\sim$ 17 \textmu{}m in diameter, roughly in the center of the waveguide, and contains $\sim3\times10^4$ NVs. We perform a Ramsey measurement on those NVs to characterize the inhomogeneous broadening. As shown in Fig.~2b, the dephasing time of the NV ensemble is $T_2^* = 0.63(2)$ \textmu{}s, corresponding to a dephasing rate $\Gamma_2^* = (2\pi\times)0.25(1)$ MHz (We omit $2\pi$ hereinafter, and the unit Hz represents the circular frequency).

We apply a magnetic field of 36.09 mT along the N-V axis of part of the NV centers, where the energy splitting between $|0\rangle$ and $|1\rangle$ is 3880.44 MHz. As a comparison, the $^{19}$F nuclear spin has an energy splitting of only 1.447 MHz. To eliminate the energy mismatch, we apply an amplitude-modulated continuous driving field $2\Omega \cos(\omega t)\cos(\omega_{\text{NV}} t)$ on the NV ensemble, where $\Omega$ is the corresponding Rabi frequency. Then the NV center has an equivalent splitting determined by the modulation frequency $\omega$, which can be easily tuned and is robust to the inhomogeneity of driving field \cite{Qin2023,Huang2025}. In such a driving field, the dephasing rate $\Gamma_2^*$, the sensor-target coupling $d$, and the RF strength are reduced by factors of $J_0(\kappa)$, $J_1(\kappa)$, and $J_1(\kappa)$, respectively (see SI Note 1), where $\kappa=\Omega/\omega$ is the relative driving index and $J_n$ is the $n$-th order Bessel function of the first kind. By sweeping $\omega$, a cross-relaxation spectrum can be obtained (Fig.~2c). The spectrum has a peak centering at 1.5 MHz, which is larger than the estimation of 1.447 MHz. The linewidth of 0.23 MHz is also larger than $J_0(2.0)4\sqrt{\ln 2}\Gamma_2^*\approx 0.19$ MHz, indicating the existence of other unresolved nuclear spins with higher Larmor frequencies. We then apply an additional RF field with frequency sweeping in sync with $\omega$. As shown in Fig.~2d, two burned holes show at 1.446 MHz and 1.534 MHz, consisting with the Larmor frequency of $^{19}$F and $^1$H nuclear spins, respectively. The latter probably comes from the adsorbed water layer on the diamond surface. When the RF is turned off, the spectrum in Fig.~2d still exhibits weak hole-burning-like characteristics at 1.446 MHz, which is induced by the relaxation of $^{19}$F nuclear spins (see SI Note 1).

\begin{figure}[htbp]
	\centering \includegraphics[width=1\columnwidth]{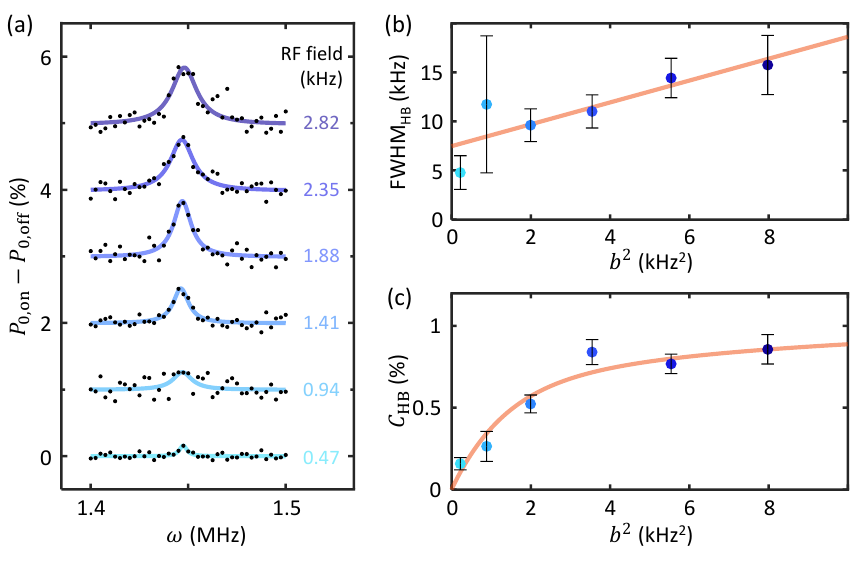} \protect\caption{
		{Performance of hole burning at different pump strength.}
		(a) Hole-burning spectra with different strength of RF fields. The data points are given by the difference between two cross-relaxation measurements with RF on and off. Meaningless offsets are added for ease of comparison. The solid lines are Lorentz fits. 
		(b) Dependence of hole-burning linewidth on RF strength. The points are fitted FWHM with error bars given by the fitting errors. The solid line is a linear fit with intercept and slope of 7(2) kHz and 1.1(5) kHz$^{-1}$, respectively. 
		(c) Dependence of hole-burning contrast on RF strength. The points are fitted peak heights with error bars given by the fitting errors. The solid line is a fit according to Eq.~8, which gives a saturation contrast of 1.2(2)\% and rate of 0.8(3) kHz$^{-2}$.
	}
	\label{detail}
\end{figure}

We further investigate the dependence of hole linewidth and contrast on the pump RF field. The RF strength is calibrated independently using the XY8 sequence (see SI Note 2). As shown in Fig.~3a, we measure a series of hole-burning spectra with different RF strength. Here we focus on the $^{19}$F nuclear spin because it has a narrower linewidth than the protons. To extract the hole linewidth and contrast, we use a Lorentz function to fit the hole-burning spectrum. As predicted by Eq.~9, the measured hole linewidth indeed shows linear dependence on $b^2$ (Fig.~3b). The fitted slope of 1.1(5) kHz$^{-1}$ roughly consists with $(2\pi\times)0.6t\approx0.75$. The fitted intercept of 7(2) kHz gives the upper bond of the decoherence rate $\Gamma_2 < 1.8(5)$ kHz. The measured hole contrast shows a saturation trend (Fig.~3c), which is well predicted by Eq.~8. Since $C_{\text{HB}} \to C_{\text{T}}$ when $b \to \infty$, the fitted saturated contrast of 1.2(2)\% is just the contrast of $^{19}$F nuclear spins in the bare cross-relaxation spectrum (Fig.~2c). It means that the signal loss in hole-burning spectroscopy can be reduced to virtually zero. The fitted saturation rate $t/\Gamma_2\sim$ 0.8(3) kHz$^{-2}$ also gives an estimated decoherence rate of $\Gamma_2 \sim 1.4(5)$ kHz, consisting with the above analysis of linewidth.

In conclusion, we give an easy way to remove the inhomogeneous broadening in cross-relaxation spectra. Except for an additional weak pump field, the measurement is almost identical to an ordinary relaxation measurement. So this method can eliminate the drawbacks of NV relaxometry in terms of spectral resolution while maintaining its convenience. Besides, the experimental results show that the spectral line resolution can be improved with a negligible degradation on measurement efficiency by choosing a proper strength of pump field. So it can be as efficient as the dynamical decoupling method (see SI note 3). Moreover, the experimental data are in good agreement with the simple theoretical model we proposed, enabling our method to be directly applied to relaxometry based on other color centers in SiC or hBN. 

The spectral resolution of the hole-burning spectra are currently comparable with the ordinary dynamical decoupling method ($\sim 1/T_2$ \cite{Staudacher2013}), but still poorer than the correlation ($\sim 1/T_1$ \cite{Laraoui2013}) and heterodyne ($\sim 1/t$ \cite{Schmitt2017,Boss2017}, $t$ the total measurement time) method based on dynamical decoupling. A prior work has shown that the NV relaxometry has also achieved $1/t$ scaling of spectral resolution when sensing AC fields with fixed phase \cite{Wang2022}. The target spin can generate such fields if it is polarized \cite{Glenn2018}. Since the cross-relaxation measurement can be accompanied by polarization of the target spin \cite{London2013}, its spectral resolution is possible to be further improved to a level comparable to that of dynamical decoupling methods.

Acknowledgments: This work was supported by the National Key R\&D Program of China (Grant No. 2024YFB3212600), the National Natural Science Foundation of China (Grant No. T2125011), the CAS (Grant No. YSBR-068), Innovation Program for Quantum Science and Technology (Grant No. 2021ZD0302200 and No. 2021ZD0303204), New Cornerstone Science Foundation through the XPLORERPRIZE, and the Fundamental Research Funds for the Central Universities. 

\renewcommand\refname{Reference}

\bibliography{HoleBurning}

\bibliographystyle{naturemag}

\clearpage
\newpage
\onecolumngrid

\setcounter{section}{0}
\setcounter{figure}{0}
\setcounter{table}{0}
\setcounter{equation}{0}

\renewcommand{\thesection}{S\arabic{section}}
\renewcommand{\thefigure}{S\arabic{figure}}
\renewcommand{\thetable}{S\arabic{table}}
\renewcommand{\theequation}{S\arabic{equation}}


\section*{Supplementary Materials}
	
\subsection*{Supplementary Note 1. Brief description of the cross-relaxation model}

The NV center has a spin-triplet ground state consisting of $|0\rangle$ and $|\pm 1\rangle$. By applying a magnetic field along the N-V axis, the degeneracy between $|1\rangle$ and $|-1\rangle$ can be lifted. If only two energy levels are involved in the measurement, the NV center can be simplified as a spin-1/2 system. For sensing of spin-1/2 nuclear spin, the Hamiltonian of the sensor-target system can be written as
\begin{equation}
	H = (\omega+\Delta) S_z + d_{ij}S_i I_j + \nu I_z + 2\Omega \cos(ft)\cos(\omega t)S_x,
\end{equation}
where $S$ and $I$ is the spin operator of the sensor and the target, respectively, $\omega$ is the energy splitting of the sensor with inhomogeneous noise $\Delta$, $\nu$ is the energy splitting of the target, $d$ is the dipole-dipole coupling between them, and $\Omega$ is the driving strength with modulation frequency $f$. In the interaction picture, the Hamiltonian becomes
\begin{equation}
	\begin{split}
	H_{\text{I}} & = e^{i\omega t S_z}He^{i\omega t S_z}-\omega S_z \\
	& \approx \Delta S_z + d_{zj}S_z I_j + \nu I_z + \Omega \cos(ft)S_x,
	\end{split}
\end{equation}
where the items with frequency $\geq\omega$ are neglected. In the second interaction picture, the Hamiltonian becomes
\begin{equation}
	\begin{split}
		H_{\text{II}} & = e^{i[\frac{\Omega}{f} \sin (ft) S_x + (ft+\varphi) I_z]}H_{\text{I}} e^{-i[\frac{\Omega}{f} \sin (ft) S_x + (ft+\varphi) I_z]}-\Omega \cos (ft) S_x - f I_z \\
		& =\Delta\left\{\sum_{-\infty}^{\infty}J_{2n}(\kappa)\cos(2nft)S_z + \sum_{-\infty}^{\infty}J_{2n+1}(\kappa)\sin[(2n+1)ft]S_y\right\} + (\nu-f)I_z \\
		& \quad + \left\{\sum_{-\infty}^{\infty}J_{2n}(\kappa)\cos(2nft)S_z + \sum_{-\infty}^{\infty}J_{2n+1}(\kappa)\sin[(2n+1)ft]S_y\right\}\cdot\\
		& \qquad \left\{d_{zx}[I_x \cos(ft+\varphi)-I_y\sin(ft+\varphi)]+d_{zy}[I_y \cos(ft+\varphi)+I_x\sin(ft+\varphi)]+d_{zz}I_z \right\} \\
		& \approx J_0(\kappa)\Delta S_z - d_{z\perp}J_1(\kappa)S_y I_y + (\nu-f)I_z,
	\end{split}
	\label{HAM}
\end{equation}
where $J_m$ is the $m$-th order Bessel functions of the first kind, $\kappa=\Omega/f$ is the relative driving index, $d_{z\perp}=\sqrt{d_{zx}^2+d_{zy}^2}$ is the lateral coupling strength, and $\varphi = \arctan(d_{zy}/d_{zx})$. Here we use the Jacobi-Anger expansion and neglect items with frequency $\geq f$. The longitudinal coupling $d_{zz}J_0(\kappa)S_z I_z$ is also neglected because it does not contribute to the relaxation. The Hamiltonian can be divided into two parts $H_{\text{II}}=H_1+H_2$, where
\begin{equation}
	\begin{split}
		& H_1 = \frac{J_0(\kappa)\Delta +\nu-f}{2}(S_z + I_z) + \frac{d_{z\perp}J_1(\kappa)}{4}(S_{+}I_{+}+S_{-}I_{-}) \\
		& H_2 = \frac{J_0(\kappa)\Delta -\nu+f}{2}(S_z - I_z) - \frac{d_{z\perp}J_1(\kappa)}{4}(S_{+}I_{-}+S_{-}I_{+}).
	\end{split}
\end{equation}
The first and second parts induce transitions of $|1,\uparrow\rangle \leftrightarrow |0,\downarrow\rangle$ and $|1,\downarrow\rangle \leftrightarrow |0,\uparrow\rangle$, respectively, with relaxation rate
\begin{equation}
	\Gamma_{\text{d1,d2}}=\frac{d_{z\perp}^2 J_1^2(\kappa) \Gamma_2}{4[\Gamma_2^2 + (\nu-f\pm J_0(\kappa)\Delta)^2]}.
\end{equation}

If the target spin does not change its state during each round of measurement, the NV population can be calculated as
\begin{equation}
	\begin{split}	
	P_0 & = \frac{1}{2}\int\frac{1}{2}\left[1+e^{-(\Gamma_1+\Gamma_{\text{d1}}) t}\right]\cdot G(\Delta)d\Delta + \frac{1}{2}\int\frac{1}{2}\left[1+e^{-(\Gamma_1+\Gamma_{\text{d2}}) t}\right]\cdot G(\Delta)d\Delta \\
	& = \int\frac{1}{2}\left[1+e^{-(\Gamma_1+\Gamma_{\text{d1}}) t}\right]\cdot G(\Delta)d\Delta = \int\frac{1}{2}\left[1+e^{-(\Gamma_1+\Gamma_{\text{d2}}) t}\right]\cdot G(\Delta)d\Delta,
	\end{split}
	\label{P0}
\end{equation}
where the noise is symmetrically distributed around zero and characterized by a Gaussian function
\begin{equation}
	G(\Delta;\sqrt{2}\Gamma_2^*) = \frac{1}{2\sqrt{\pi}\Gamma_2^*}e^{-\frac{\Delta^2}{4\Gamma_2^{*2}}}.
\end{equation}
For $\xi=\Gamma_{\text{d,max}}t=d_{z\perp}^2 J_1^2(\kappa) t/(4\Gamma_2) \ll 1$, the exponential Lorentz function $\exp[-\Gamma_{\text{d}}t]$ can be approximated by a Lorentz function $1-\Gamma_{\text{d}}t$. As shown in Fig.~S1, this approximation can be generalized to a larger scale (such as $\xi \leq 5$) as 
\begin{equation}
	e^{-\frac{\xi}{1 + x^2}} \approx 1-\frac{(1-e^{-\xi})\alpha^2}{\alpha^2+x^2},
	\label{expL}
\end{equation}
where
\begin{equation}
	\alpha = \sqrt{\frac{\xi}{\ln 2-\ln(1+e^{-\xi})}-1}\approx 1+0.3\xi.
\end{equation}
Substituting this approximation into Eq.~\ref{P0}, we have
\begin{equation}
	\begin{split}
		P_0 & \approx \frac{1}{2}+\frac{e^{-\Gamma_1 t}}{2} \int\left[1-\frac{(1-e^{-\xi})(\alpha \Gamma_2)^2}{(\alpha\Gamma_2)^2 + (\nu-f+ J_0\Delta)^2}\right]G(\Delta)d\Delta \\
		& = \frac{1}{2}+\frac{e^{-\Gamma_1 t}}{2}-\frac{\pi\alpha\Gamma_2 e^{-\Gamma_1 t}(1-e^{-\xi})}{2} \int\frac{\alpha \Gamma_2}{\pi[(\alpha\Gamma_2)^2 + (\nu-f+ J_0\Delta)^2]} \cdot \frac{1}{2\sqrt{\pi}J_0\Gamma_2^*}e^{-\frac{( J_0\Delta)^2}{4(J_0\Gamma_2^*)^2}}d( J_0\Delta) \\
		& = \frac{1}{2}+\frac{e^{-\Gamma_1 t}}{2}-\frac{\pi\alpha\Gamma_2 e^{-\Gamma_1 t}(1-e^{-\xi})}{2} L(f-\nu;\alpha\Gamma_2)*G(f;\sqrt{2}J_0\Gamma_2^*) \\
		& \approx \frac{1}{2}+\frac{e^{-\Gamma_1 t}}{2}-\frac{\pi\alpha\Gamma_2 e^{-\Gamma_1 t}(1-e^{-\xi})}{2} G(f-\nu;\sqrt{2}J_0\Gamma_2^*) \\
		& = \frac{1}{2}+\frac{e^{-\Gamma_1 t}}{2}-\frac{\sqrt{\pi}\alpha\Gamma_2 e^{-\Gamma_1 t}(1-e^{-\xi})}{4J_0\Gamma_2^*} e^{-\frac{(f-\nu)^2}{4(J_0\Gamma_2^*)^2}},
	\end{split}
	\label{P0spectrum}
\end{equation}
where the second approximation is made by replacing the Lorentz function with the delta function if $\alpha\Gamma_2 \ll J_0\Gamma_2^*$.

\begin{figure}[h]
	\centering
	\includegraphics[width=0.9\columnwidth]{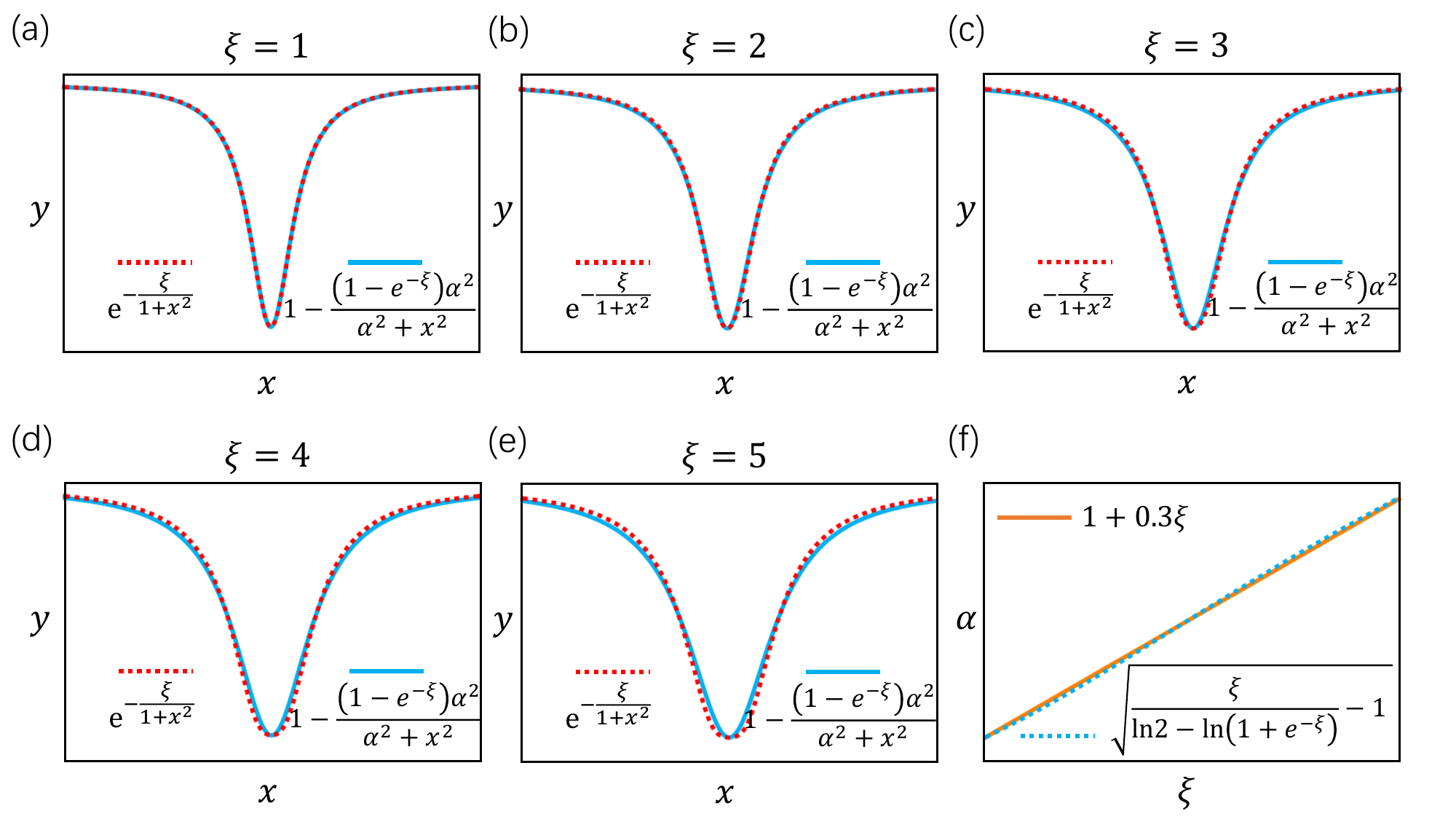} \caption{\textbf{Approximation of exponential Lorentz function.}
		(a-e) Comparison between a exponential Lorentz function $\exp[-\xi/(1+x^2)]$ and a broadened Lorentz function $1-[1-\exp(-\xi)]\alpha^2/(\alpha^2+x^2)$ with different $\xi$. 
		(f) Dependence of the broadening factor $\alpha$ on $\xi$.  
	}
	\label{singleND}
\end{figure}

For the extreme case, the target spin flips fast enough that the relaxation rate averages out to
\begin{equation}
	\langle \Gamma_{\text{d}} \rangle = \frac{\Gamma_{\text{d1}}+\Gamma_{\text{d2}}}{2}.
\end{equation}
Since $e^{-\Gamma_{\text{d}}t}$ is a concave function with respect to $\Gamma_{\text{d}}$, we have
\begin{equation}
	e^{-\langle \Gamma_{\text{d}} \rangle t}\leq \frac{e^{-\Gamma_{\text{d1}}t}+e^{-\Gamma_{\text{d2}}t}}{2},
\end{equation}
with equality at $f=\nu$. In reality, the target spin will change its state at a moderate rate, leading to a partially averaged relaxation rate. Even so, Eq.~\ref{P0spectrum} is still an overestimation of the NV population except for $f=\nu$. That is why there exists a peak at $f=\nu$ in Fig.~2d in the main text.

\subsection{Supplementary Note 2. Calibration of the RF field strength}
We use the XY8 sequence to measure the z-directed strength of the RF field, as shown in Fig.~S2(a). Knowing that the RF field strength $b$ is directly proportional to the RF output voltage $u$ with a proportionality constant of $k$, i.e., $b = k\cdot u$, the accumulated phase during the XY8-N sequence should be
\begin{equation}
	\phi_\textrm{RF} = N\cdot \int_0^{\pi/f_\textrm{RF}} b \sin(f_\textrm{RF} t) d t = \frac{2k\cdot u\cdot N}{f_\textrm{RF}}.
	\label{RF}
\end{equation}
By sweeping $u$, the readout NV population $P_0 = 1/2 +  1/2 \cos(\phi_\textrm{RF})$ will oscillate with frequency of $\nu_\textrm{RF} = 2kN/f_\textrm{RF}$ (Fig.~S2(b)). Then the proportionality constant $k$ is determined by
\begin{equation}
	k = \frac{\nu_\textrm{RF} f_\textrm{RF}}{2 N}.
\end{equation}
For the RF strength in the main text, it needs to be multiplied by $J_1(\kappa)$.

\begin{figure}[h]
	\centering \includegraphics[width=0.7\columnwidth]{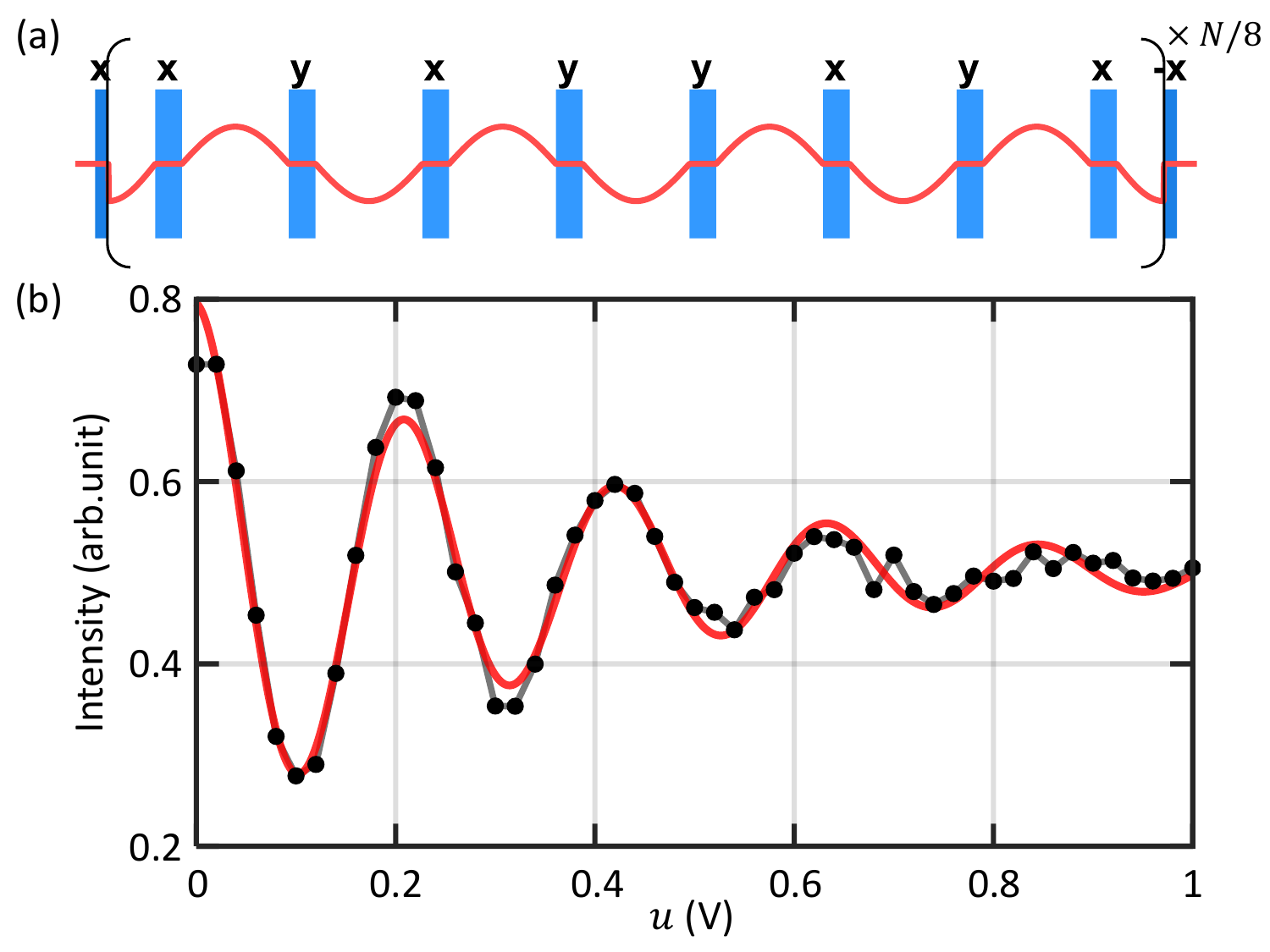} \protect\caption{
		\textbf{RF strength calibration on NV center.}
		(a) XY8-N sequence for sensing of RF fields. The interval between $\pi$ pulses is equal to half of the RF period. Here the RF is intentionally set as segmented to skip $\pi$ pulses.
		(b) Measurement of RF fields. The points are experimental results, while the line is a cosine damping fit, giving $\nu_\textrm{RF} = 4.71(3)$ V$^{-1}$.
	}
\end{figure}

\subsection{Supplementary Note 3. Sensitivity of the cross-relaxation measurement}
The dynamical decoupling (DD) measurement of nuclear spins is similar to that of RF fields described above, where the field generated by the nuclear spin has a random phase. Similar to Eq.~\ref{RF}, the signal contrast of DD measurement in the small-signal limit is
\begin{equation}
	C_{\text{DD}} = \frac{\langle \phi^2\rangle}{4} e^{-(t/T_2)^2} = \frac{b^2 N^2 \tau^2}{2\pi^2} e^{-(t/T_2)^2} = \frac{d_{z\perp}^2 t^2}{8\pi^2} e^{-(t/T_2)^2},
\end{equation}
where $b=d_{z\perp}/2$ is the sensor-target coupling strength, $\tau=\pi/\nu$ is the interval between $\pi$ pulses, and $t$ is the evolution time. The signal-to-noise ratio (SNR) with a total measurement time of $T_{\text{tot}}$ is
\begin{equation}
	\text{SNR}_{\text{DD}} = C_{\text{DD}}\sqrt{N_0\frac{T_{\text{tot}}}{t}} \leq \frac{1}{8 \pi^2}\left(\frac{3}{4e}\right)^{\frac{3}{4}} \sqrt{N_0 T_{\text{tot}}} d_{z\perp}^2 T_2^{\frac{3}{2}},
\end{equation}
where the optimal $t=\sqrt{3}T_2/2$, and $N_0$ is the number of collected photons in a single round of measurement. According to Eq.~\ref{P0spectrum}, the signal contrast of the amplitude-modulated (AM) cross-relaxation spectrum in the small-signal limit is
\begin{equation}
	C_{\text{AM}} = \frac{\sqrt{\pi}\Gamma_2 e^{-\Gamma_1 t}}{4J_0(\kappa) \Gamma_2^*}\cdot \frac{d_{z\perp}^2 J_1^2(\kappa) t}{4\Gamma_2}.
\end{equation}
The SNR with a total measurement time of $T_{\text{tot}}$ is
\begin{equation}
	\text{SNR}_{\text{AM}} = C_{\text{AM}}\sqrt{N_0\frac{T_{\text{tot}}}{t}} \leq \frac{\sqrt{\pi}J_1^2(\kappa)}{16J_0(\kappa)}\left(\frac{1}{2e}\right)^{\frac{1}{2}} \sqrt{N_0 T_{\text{tot}}} d_{z\perp}^2 T_2^* \sqrt{T_1^{'}},
\end{equation}
where the optimal $t=T_1^{'}/2$. $T_1^{'}$ is the effective relaxation time under amplitude-modulated driving, which is shorter than the normal relaxation time $T_1$ (Fig.~S3(a)), and positively depends on the modulated frequency $\omega$ (Fig.~S3(b)). The comparison between the two kind of measurements is
\begin{equation}
	\frac{\text{SNR}_{\text{AM}}}{\text{SNR}_{\text{DD}}} \approx 10\times\frac{J_1^2(\kappa)}{J_0(\kappa)} \cdot \frac{T_2^* \sqrt{T_1^{'}}}{T_2^{\frac{3}{2}}}.
\end{equation}
Although $T_1^{'}$ negatively depends on $\kappa$ (Fig.~S3(c)), this ratio still increases with increasing $\kappa$ until $\kappa \sim 2.4$, at which point this formula becomes invalid due to divergence. In fact, at this point, the inhomogeneous broadening can no longer be estimated by $J_0(\kappa) \Gamma_2^*$, as other higher-order effects in Eq.~\ref{HAM} will emerge. We also experimentally compare the sensitivity of the two kind of measurements. As shown in Fig.~S3(d)(e), the experimentally measured SNR ratio $\text{SNR}_{\text{AM}}/\text{SNR}_{\text{DD}}=32/12\times\sqrt{8.3/85.3}\approx0.83$. Therefore, the efficiency of cross-relaxation measurement is comparable with that of dynamical decoupling.

\begin{figure}[h]
	\centering \includegraphics[width=0.9\columnwidth]{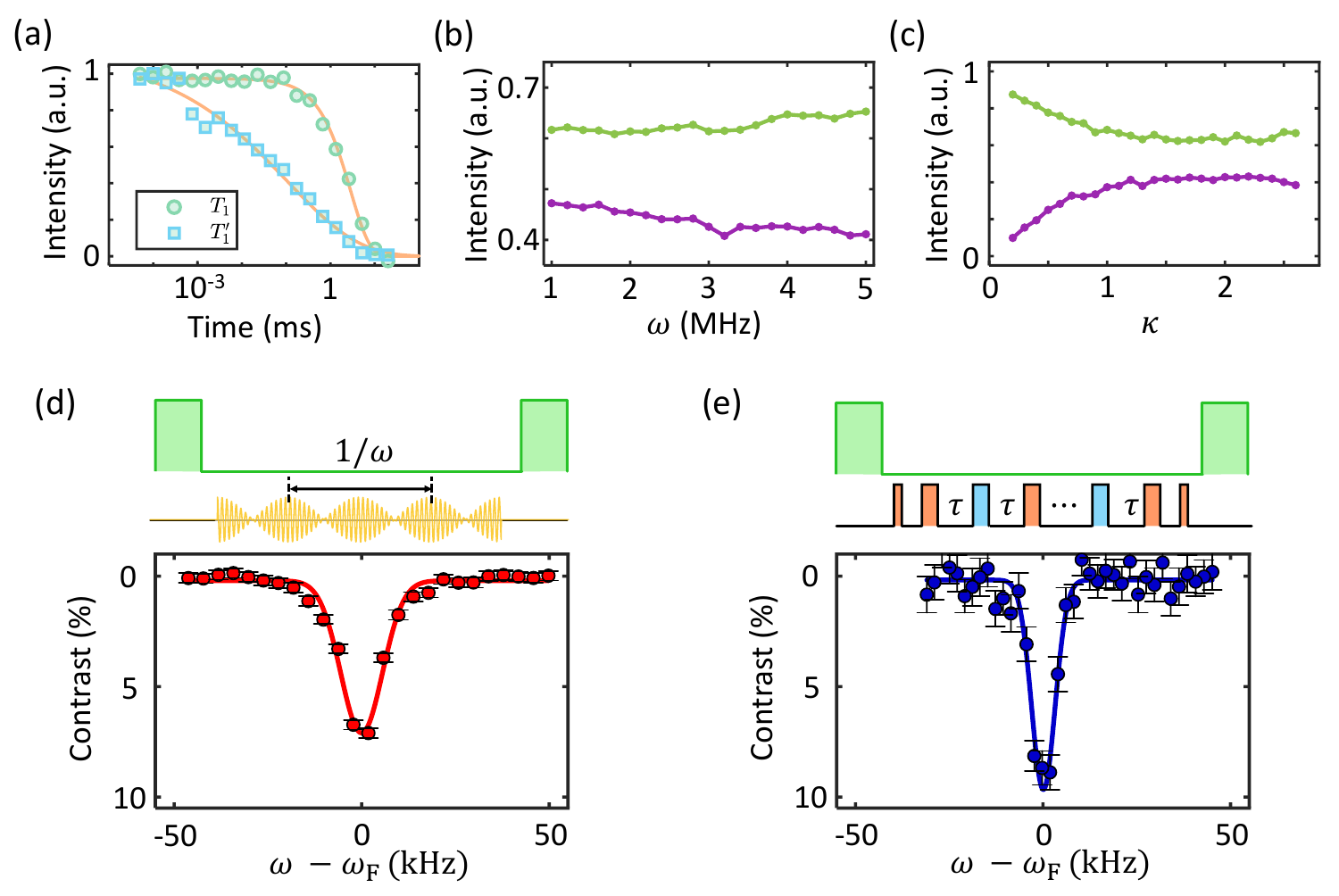} \protect\caption{
		\textbf{Sensitivity of cross-relaxation measurement.}
		(a) Relaxation under amplitude-modulated driving. The points are experimental results, while the lines are exponential fits giving $T_1=3.0(1)$ ms, $T_1^{'}=70(20)$ \textmu{}s. The driving strength and modulation frequency are 3 MHz and 1.4 MHz, respectively.
		(b) Dependence of relaxation on modulated frequency $\omega$. The upper and lower sets of data correspond to initial states of $|0\rangle$ and $| 1 \rangle$, respectively. The evolution time and $\kappa$ are fixed at 1 ms and 1.8, respectively.
		(c) Dependence of relaxation on relative driving index $\kappa$. The evolution time and $\omega$ are fixed at 1 ms and 4 MHz, respectively.
		(d) Amplitude-modulated cross-relaxation measurement of Fomblin oil. The evolution time is 200 \textmu{}s. $\kappa = 2.4$. The SNR of the resonance peak is 32 with single-frequency-point time consumption $T_{\text{tot}} = 85.3$ s.
		(e) Dynamical decoupling (XY8-512) measurement of Fomblin oil. The evolution time is $\sim512*0.34\approx174$ \textmu{}s. The SNR of the resonance peak is 12 with single-frequency-point time consumption $T_{\text{tot}} = 8.3$ s. 
	}
\end{figure}

\end{document}